\documentclass[aps,prx,twocolumn,floatfix,superscriptaddress]{revtex4}

\usepackage{amssymb,amsmath,amstext}                
\usepackage{graphicx}                                               
\usepackage{epstopdf}                                               
\usepackage{color}                                                     
\usepackage{bm}                                                        
\usepackage{appendix}                                              
\usepackage[utf8]{inputenc}
\usepackage{bbold}
\usepackage{bbm}
\usepackage{ulem}
\usepackage{latexsym}
\usepackage[colorlinks=true,citecolor=blue,linkcolor=magenta]{hyperref}

\newcommand{\vect}[1]{\mathbf{#1}}

\def\be{\begin{equation}}
\def\ee{\end{equation}}
\def\bea{\begin{eqnarray}}
\def\eea{\end{eqnarray}}
\def\ra{\rangle}
\def\la{\langle}
\def\bi{\begin{itemize}}
\def\ei{\end{itemize}}
\def\ben{\begin{enumerate}}
\def\een{\end{enumerate}}

\begin{document} 

\title{Anderson Molecules}

\author{Pawe\l{} Matus} 
\affiliation{
Instytut Fizyki Teoretycznej, 
Uniwersytet Jagiello\'nski, ulica Profesora Stanis\l{}awa \L{}ojasiewicza 11, PL-30-348 Krak\'ow, Poland}
\author{Krzysztof Giergiel} 
\affiliation{
Instytut Fizyki Teoretycznej, 
Uniwersytet Jagiello\'nski, ulica Profesora Stanis\l{}awa \L{}ojasiewicza 11, PL-30-348 Krak\'ow, Poland}
\author{Krzysztof Sacha} 
\affiliation{
Instytut Fizyki Teoretycznej, 
Uniwersytet Jagiello\'nski, ulica Profesora Stanis\l{}awa \L{}ojasiewicza 11, PL-30-348 Krak\'ow, Poland}

\begin{abstract}
Atoms can form molecules if they attract each other. Here, we show that atoms are also able to form bound states not due to the attractive interaction but because of destructive interference. If the interaction potential changes in a disordered way with a change of the distance between two atoms, Anderson localization can lead to the formation of exponentially localized bound states. While disordered interaction potentials do not exist in nature, we show that they can be created by means of random modulation in time of the strength of the original interaction potential between atoms and objects that we dub Anderson molecules can be realized in the laboratory.
\end{abstract}

\date{\today}

\maketitle

\section{Introduction}

If the interaction potential between two particles depends on  distance between them and possesses sufficiently large potential well, particles can form a bound state where their relative distance is fixed. Interactions between two atoms can be described by the van der Waals forces which include long-range attraction  and short-range repulsion. The resulting potential well can be deep enough to support bound states of atoms \cite{Blaney1976}. This is an example of molecules which can form provided there are attractive interactions between particles. 

Let us imagine that the interaction potential between two atoms changes in a disordered way as a function of their relative distance. It means that the degree of freedom corresponding to the relative position is described by a similar Hamiltonian as the Hamiltonian for a single particle moving in a disordered potential. In the latter case it is well known that Anderson localization is possible where eigenstates of a particle are exponentially localized in configuration space \cite{Anderson1958,MuellerDelande:Houches:2009}. In the case of two atoms interacting via a disordered potential, the Anderson localization would mean that the atoms form a bound state where their relative distance is defined with uncertainty given by the Anderson localization length. Formation of this kind of bound states would be a result of destructive interference between different multiple scattering paths similarly as in the standard Anderson localization case \cite{MuellerDelande:Houches:2009}. Although the idea may sound simple, disordered interaction potentials cannot be found in nature. However, they can be created by means of {\it time engineering} which has been used in the field of  time crystals to realize different condensed matter phases in the time domain \cite{Sacha2017rev,Guo2020}.

When a periodically moving particle is resonantly driven by periodically changing external force, its motion in the frame moving along an unperturbed periodic orbit can be described by an effective solid state-like Hamiltonian, i.e. the effective Hamiltonian is similar like for an electron moving in a crystalline potential formed by ions \cite{Guo2013,Sacha15a}. Importantly, when such a particle is observed in the laboratory frame, solid state behavior is revealed in the time domain \cite{Sacha15a,sacha16}. It turns out that the potential in the effective Hamiltonian can be engineered nearly at will by means of a proper choice of Fourier components of the periodically changing external force and single-particle and many-body condensed matter phenomena, ranging from Anderson localization, topological crystals and quasi-crystals to  Mott-insulator phase and many-body localization, can be observed in the time domain \cite{Guo2013,Sacha15a,sacha16,Guo2016,Guo2016a,Giergiel2017,
delande17,
Mierzejewski2017,Liang2017,Giergiel2018,Giergiel2018a,
Lustig2018,Giergiel2018b}. This is the field of condensed matter physics in time crystals (for reviews see \cite{Sacha2017rev,Guo2020}). It should be stressed that the time crystal research concerns also investigation of systems that are able to break spontaneously time translation symmetry \cite{Wilczek2012,Bruno2013b,Watanabe2015,Kozin2019,
Ohberg2019,SyrwidKosiorSacha2020,
ReplyOhbergWright2020,Syrwid2020,You2020,Sacha2017rev}. For periodically driven systems it means that the discrete time translation symmetry, dictated by an external drive, is spontaneously broken and the so-called discrete time crystals can form \cite{Sacha2015,Khemani16,ElseFTC,Huang2017,
Russomanno2017,Gong2017,Iemini2017,Kosior2018,Kosior2018a,
Giergiel2018c,Surace2018,Matus2019,Gambetta2018,Zhu2019,
Pizzi2019,Pizzi2019a,khemani2019brief,Cai2020,
Giergiel2020,Kuros2020,Lazarides2019} which has been already demonstrated experimentally \cite{Zhang2017,
Choi2017,Pal2018,Rovny2018,Smits2018}.

In the present paper we show that not only an external potential for atoms can be engineered by means of a proper time modulation of an external force but also an effective interaction potential between atoms can be controlled and engineered if atomic s-wave scattering length is modulated in time. It allows one to create an effective interaction potential that changes in a disordered way as a function of the distance between atoms and bound states (which we dub Anderson molecules) can be realized in the laboratory. This idea was briefly mentioned in our previous publication \cite{Giergiel2018}.

The paper is organized as follows. In Sec.~\ref{sec1} we introduce the basic idea of Anderson molecules by considering two atoms moving on a one-dimensional (1D) ring. In Sec.~\ref{sec2} it is shown how to create a pair of Anderson molecules and how to control interactions between them. Section~\ref{sec3} is devoted to analysis of the formation of Anderson molecules by atoms moving in a box potential which seems to be easier to realize in the laboratory, especially in the 3D case. The latter is described in Sec~\ref{sec4}. Summary is given in Sec~\ref{conclusions} and some more technical information is provided in Appendixes~\ref{appendixA}-\ref{appendixC}.


\begin{figure} 	            
\includegraphics[width=1.0\columnwidth]{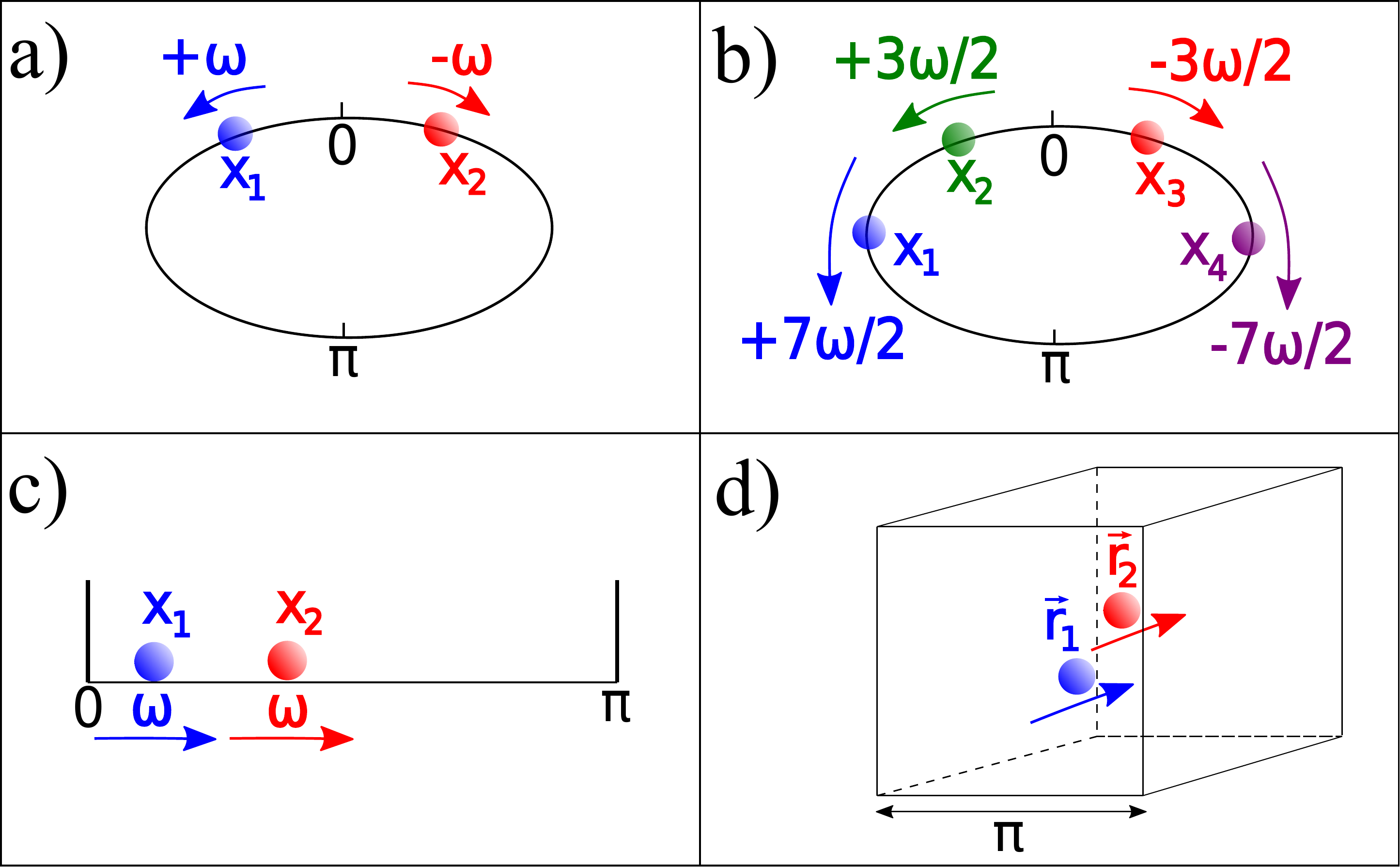}       
\caption{Four panels illustrate four different experimental setups for the realization of Anderson molecules which are considered in the present paper. Panel~(a): two atoms are moving in the opposite directions on a ring with the velocities $\pm \omega$. If the strength of the interaction between atoms is modulated in time in a disordered way, a diatomic Anderson molecule can form. This is the basic experimental setup which is described in Sec.~\ref{sec1}. Panel~(b): if four atoms are moving on a ring with the velocities $\pm3\omega/2$ and $\pm7\omega/2$ and the strengths of the mutual interactions are properly modulated in time, two Anderson molecules form and the interaction potential between them can be controlled experimentally, see Sec.~\ref{sec2}. Panel~(c): two atoms which move initially with the velocity $\omega$ in a 1D potential well can also form an Anderson molecule and its experimental detection is analyzed in Sec.~\ref{sec3}. Panel~(d) shows two atoms which move initially with the same velocity in a 3D potential well. This setup allows one to realize an Anderson molecule in 3D space which is described in Sec.~\ref{sec4}.
}
\label{all_setups}   
\end{figure} 

\section{Anderson molecule on a ring}
\label{sec1}

At low energies collisions of two atoms are characterized by one parameter only, i.e. the s-wave scattering length. Consequently, interactions between atoms can be modeled by any potential provided it reproduces the correct value of the scattering length \cite{Castin_LesHouches}. In ultra-cold atoms the interactions are usually modeled by means of the zero-range potential proportional to the Dirac-delta function. The strength of the potential is determined by the s-wave scattering length which can be controlled experimentally  by means of a magnetically tunnable Feshbach resonance which occurs when two colliding atoms resonantly couple to a molecular bound state \cite{Chin2010}. In the present paper, except Sec.~{\ref{sec4}}, we assume that atoms are confined in the quasi-1D space. That is, there is a trapping potential along the two transverse directions which is so strong that excited states corresponding to the transverse degrees of freedom are not attainable for ultra-cold atoms and atoms behave like a 1D system. In the quasi-1D case, coupling between two atoms and a transversally excited molecular bound states can lead to the so-called confinement-induced resonances which provide another method for changes and control of the interaction strength between atoms \cite{Olshanii1998}.

Let us begin with two distinguishable atoms on a 1D ring of radius $R$ (similar analysis as we describe in the present paper can also be carried out for indistinguishable atoms). We assume the same kind of atoms with the mass $m$ but in different hyperfine states. In the units $\hbar^2/mR^2$ and $R$ for energy and length, respectively, the Hamiltonian of the system reads,
\begin{equation}
H = \frac{p_1^2 + p_2^2}{2} + 2\pi[\lambda_0+\lambda f(t)]\delta(x_1-x_2), 
\label{hamilt_ring_lab}
\end{equation}
where $x_i$ and $p_i$ are positions of atoms on a ring and their conjugate momenta and $\lambda_0=mR\omega_\perp a_s/(\pi \hbar)$ where $a_s$ is the atomic s-wave scattering length and $\omega_\perp$ is the frequency of the harmonic trapping potential along the transverse directions \cite{Pethick2002,Giergiel2018a}. We assume that the s-wave scattering length is periodically modulated in time using a Feshbach resonance or a confinement-induced resonance. The parameter $\lambda$ characterizes the amplitude of the modulation and 
\begin{equation}
f(t+T)=f(t) = \sum_{k=-k_m}^{k_m} f_k e^{i k \omega t},
\label{f_coeffs}
\end{equation}
where $T=2\pi/\omega$ and the coefficients $f_k$ are complex numbers which satisfy the equality $f_{-k} = f_k^*$ and $f_0=0$. 

Suppose that the first atom is moving on a ring with the velocity $\omega$ and the other one with $-\omega$, see Fig. \ref{all_setups}(a). In other words the period of the motion of the atoms along the ring is close to the period $T$ of the temporal modulation of the s-wave scattering length. The contact interaction potential that we use to model the atom-atom interactions is valid if the relative momentum of colliding atoms is much smaller than the inverse of the range $r_0$ of the true interaction potential multiplied by $\hbar$ \cite{Dalibard98}. In the units we use it means $\omega\ll R/r_0$. 

The motion of the atoms with the velocities $\pm\omega$ is resonant to the periodic modulation of the s-wave scattering length and in order to simplify the description of the system we are going to derive an effective secular Hamiltonian \cite{Berman1977,Lichtenberg1992,Buchleitner2002}. First let us switch to the frame of reference co-moving with the atoms by means of the unitary transformation $H \rightarrow UHU^\dagger + i\left(\partial_t U\right)U^\dagger$, where 
\begin{equation}
U(t) = \exp[i\omega t (p_1 - p_2)]\cdot\exp[-i\omega(x_1-x_2) ],
\end{equation}
which results in
\bea
H&=&\frac{p_1^2+p_2^2}{2} + \sum_{n=-\infty}^\infty\left(\lambda_0+\lambda \sum_{k=-k_m}^{k_m}f_{k}e^{ik\omega t}\right)
\cr &&
\times e^{in(x_1-x_2+2\omega t)},
\label{exactHmov}
\eea
where a constant term was omitted and the plane wave representation of the Dirac-delta function was used, i.e. $\delta(x)=(2\pi)^{-1}\sum_n e^{in x}$. If we are interested in quantum states for which the momenta of atoms in the moving frame are close to zero, we may average the Hamiltonian (\ref{exactHmov}) over time assuming that all quantities are slowly varying (see Appendixes~\ref{appendixA}-\ref{appendixB}) which yields the desired secular Hamiltonian,
\begin{equation}
H_{\rm eff} = \frac{p_1^2+p_2^2}{2} + \lambda\sum_{n=-k_m/2}^{k_m/2} f_{-2n}e^{in(x_1-x_2)}+\lambda_0. 
\label{floquet_2atoms}
\end{equation}
One may look at this Hamiltonian from different points of view. We have obtained it as an effective Hamiltonian within the secular approximation approach \cite{Lichtenberg1992,Berman1977,Buchleitner2002}, but it is also related to the lowest-order Magnus expansion of the Floquet evolution operator \cite{Shirley1965}, see Appendixes~\ref{appendixA}-\ref{appendixC} for details. Eigenstates of $H_{\rm eff}$ are approximate Floquet states of the system in the moving frame, where the so-called micromotion is neglected \cite{Anisimovas2015}. The range of the validity of $H_{\rm eff}$ can be estimated by an analysis of the next terms in the Magnus expansion which are negligible provided
\bea
\omega &\gg& \lambda F, \; E_{sys},
\eea
and 
\bea
\omega^2&\gg& \lambda F k_m^3, \; k_m\left(\frac{\lambda_0}{\lambda F}\right)^2,\; E_{sys}\left(\frac{\lambda_0}{\lambda F}\right)^2,  \; \frac{\lambda_0^2}{E_{sys}},
\eea 
where $F = \mathrm{max}|f_{k}|$ and $E_{sys}$ is the energy of the system in the moving frame. The effective Hamiltonian (\ref{floquet_2atoms}) does not depend on $\omega$. However, the more harmonics are present in the time-periodic function (\ref{f_coeffs}), the greater $\omega$ is needed for the effective Hamiltonian $H_{\rm eff}$ to reproduce properly the resonant dynamics described by the original Hamiltonian (\ref{hamilt_ring_lab}).

With the help of the time-independent Hamiltonian (\ref{floquet_2atoms}), it is straightforward to analyze the behavior of the system. The effective interaction potential in $H_{\rm eff}$ depends on the relative distance between atoms and consequently the center of mass degree of freedom decouples from the relative coordinate. In terms of the center of mass coordinate $X =(x_1+x_2)/2$ and the relative distance coordinate $x=x_1-x_2$, as well as their canonically conjugate momenta $P=p_1+p_2$ and $p=(p_1-p_2)/2$, respectively, the effective Hamiltonian reads
\be
H_{\rm eff} = H_{CM}+H_{rel},
\label{floquet_main}
\ee
where
\bea
H_{CM}& =& \frac{P^2}{4} ,
\label{floquet_main_CM} \\
H_{rel}& =& p^2 + \lambda\sum_{n=-k_m/2}^{k_m/2} f_{-2n}e^{inx},
\label{floquet_main_rel}
\eea
and we omit the constant term.

Let us assume that the atomic s-wave scattering length is modulated in time so that $f(t)$ behaves like a random function for $t\in [0,T)$. Because $f(t)$ is a periodic function, the same random behavior is repeated every period $T$. As an example let us consider the Fourier coefficients $f_k = e^{i \phi_k}/ \sqrt{k_m}$ where the phases are randomly drawn from the uniform distribution, i.e $\phi_k \in [0,2\pi)$ for $k>0$, and $\phi_{-k} = -\phi_k$. It means that the degree of freedom corresponding to the relative distance between the atoms, Eq.~(\ref{floquet_main_rel}), behaves like a particle moving in a 1D time-independent disordered potential characterized by the standard deviation $\lambda$ and the correlation length proportional to $1/k_m$. The theory of Anderson localization \cite{Anderson1958,MuellerDelande:Houches:2009} implies that eigenstates $\psi_E(x)$ of the Hamiltonian $H_{rel}$ are localized around different values $x_*$ of the relative distance $x$ and the probability density has an approximate exponential profile $|\psi_E(x)|^2 \sim \exp\left[-|x-x_*|/l_{loc}(E)\right]$. For weak disordered potential in (\ref{floquet_main_rel}), the localization length $l_{loc}$ can be obtained within the Born approximation,
\begin{equation}
l_{loc}(E) = \frac{2k_mp^2}{\pi\lambda^2} = \frac{2 k_m E}{\pi\lambda^2},
\label{lloc}
\end{equation}
where $E$ is the eigenenergy of $H_{rel}$ \cite{Giergiel2018, MuellerDelande:Houches:2009}. The Born approximation is valid if $\lambda^2/k_m^2\ll E < k_m^2/16$. Because the motion of the atoms is restricted to the ring of the circumference $2\pi$, Anderson localization can be observed provided $l_{loc} \ll 2\pi$. 

\begin{figure} 	            
\includegraphics[width=1.\columnwidth]{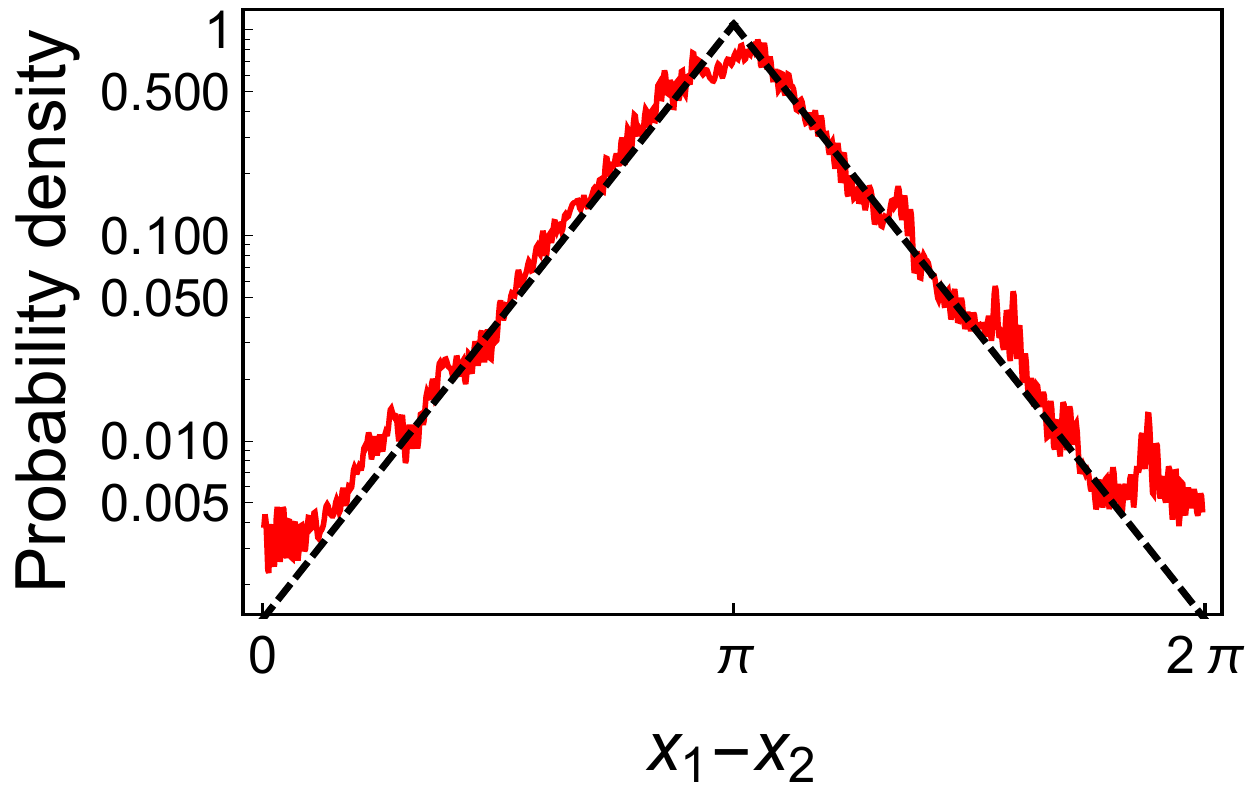}       
\caption{Red line: probability density $\overline{|\psi_E(x_1-x_2)|^2}$ corresponding to  eigenstates of the Hamiltonian (\ref{floquet_main_rel}). The overbar denotes averaging over 50 realizations of the random phases $\phi_k$ in the Fourier coefficients $f_k=e^{i\phi_k}/\sqrt{k_m}$ and over the eigenenergies in the interval $5900\le E\le5950$. The eigenstates are centered so that the probability density profile has a peak at $x_1-x_2=\pi$. Black dashed line: an approximate exponential profile $|\psi_E(x_1-x_2)|^2 \sim \exp\left[-|x_1-x_2-\pi|/l_{loc}(E)\right]$ where the localization length is given by Eq.~(\ref{lloc}) with $E=5925$. The parameters of the secular Hamiltonian (\ref{floquet_main_rel}) are the following: $\lambda = 2000$ and $k_m = 500$. The presented results are valid if $\omega \gg 10^5$.
}
\label{averaged}   
\end{figure}

Figure~\ref{averaged} shows an example of the probability density $|\psi_E(x)|^2$ averaged over many different realizations of the random phases of the coefficients $f_k = e^{i \phi_k}/ \sqrt{k_m}$ and over a small range of the eigenenergies $E$. In this example, the atoms stay at the opposite points on the ring, i.e. at the distance $x=x_1-x_2\approx x_*= \pi$, with the uncertainty determined by the localization length $l_{loc}\approx 0.47$.
It should be stressed that eigenstates $\psi_E(x)$ with similar eigenenergies $E$, and thus with similar $l_{loc}(E)$, can be localized at many different values of $x_*$. Thus, we can find eigenstates where the relative distances $x_*$ between two atoms are very different but they are characterized by similar uncertainty $l_{loc}$. 

Eigenstates $\chi(X)$ of the center of mass Hamiltonian $H_{CM}$, Eq.~(\ref{floquet_main_CM}), can be determined independently from the eigenstates of the relative position degree of freedom. They can be chosen as eigenstates of the total momentum $P$ of the atoms and consequently the probability density $|\chi(X)|^2=\rm constant$. Thus, the total eigenstates of the system, $\Psi(x,X)=\psi_E(x)\chi(X)$, show that two atoms form a bound state where their relative distance is fixed but their center of mass behaves like a free particle. The formation of such bound states is the result of the Anderson localization phenomenon and therefore we dub them {\it Anderson molecules}. 

If an eigenenergy $E$ of the relative position degree of freedom increases, the localization length $l_{loc}(E)$ increases too and at some point $l_{loc}(E)$ becomes comparable with the circumference of the ring. Then, the bound state of the atoms disappears because the uncertainty of their relative distance is comparable to the range of the entire 1D space. Hence, in the 1D case considered in the present section, dissociation of an Anderson molecule takes place gradually with an increase of its energy and the bounding energy is not sharply defined. In Sec~\ref{sec4} we describe 3D Anderson molecules where Anderson localization reveals a second order phase transition, i.e. there is a value of the energy called the mobility edge which separates localized and delocalized eigenstates. There, the bounding energy of an Anderson molecule can be related to the difference between the mobility edge and the eigenenergy $E$.

\begin{figure} 	            
\includegraphics[width=1.0\columnwidth]{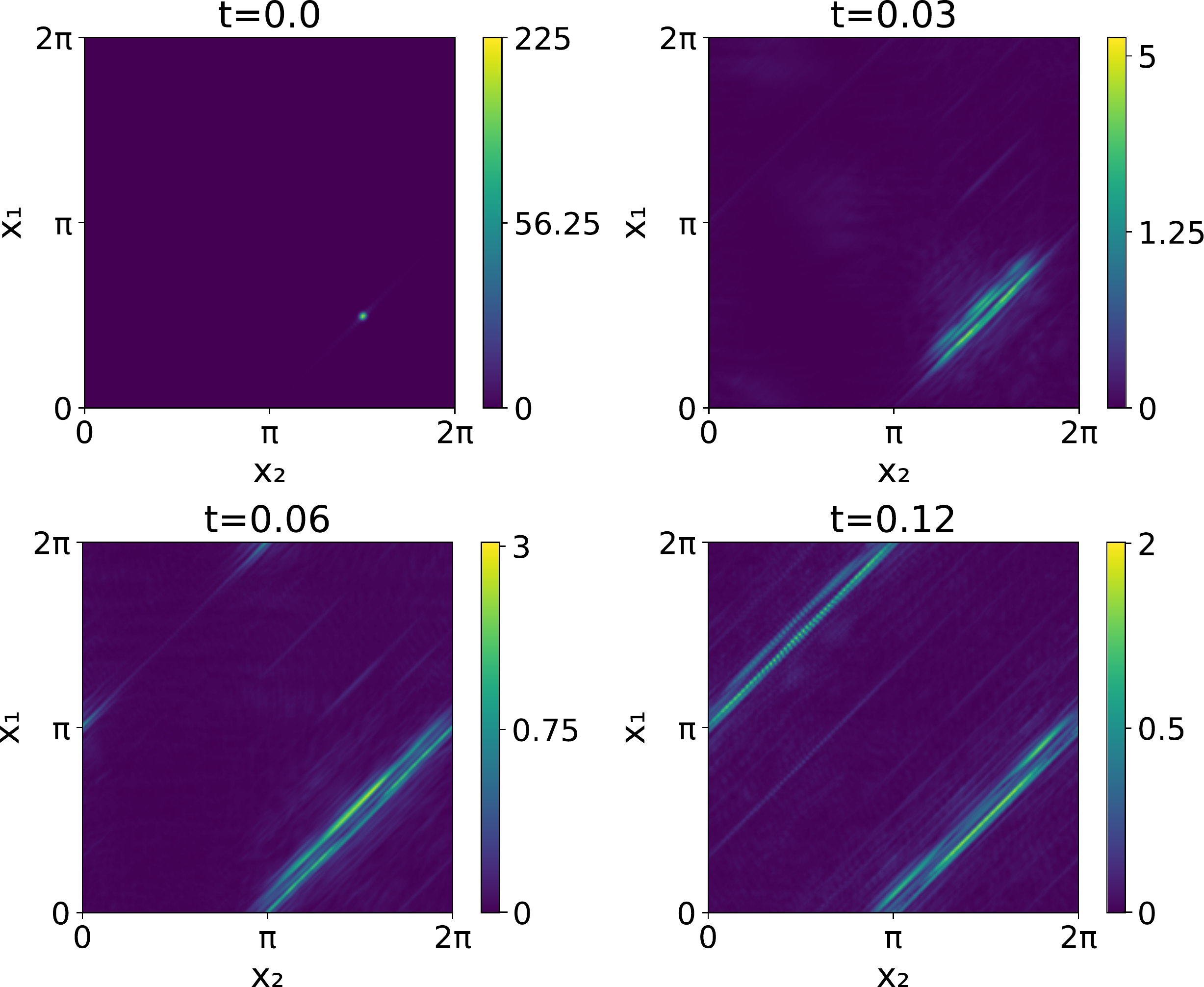}       
\caption{Time evolution of the modulus of the wavefunction $|\psi(x_1, x_2, t)|$ in the moving frame of reference. [Numbers in the color bars correspond to the values of the probability density $|\psi(x_1, x_2, t)|^2$.] The initial state is a Gaussian with the variance $\sigma^2 = 0.001$ and centered at $x_1=\pi/2$ and $x_2=3\pi/2$. The parameters of the secular Hamiltonian (\ref{floquet_main}) are $\lambda = 2000$ and $k_m = 500$. Time evolution reveals spreading of the Gaussian wavepacket along the center of mass coordinate $X=(x_1+x_2)/2$ and Anderson localization along the coordinate of the relative position of atoms $x=x_1-x_2$. At $t\approx 0.14$ the probability distribution freezes and nearly does not change in time. The results are valid if $\omega \gg 10^5$.}
\label{evolution1}   
\end{figure}

Let us discuss how Anderson molecules can be realized and detected in the laboratory. Assume that at $t=0$ one atom is prepared in a Gaussian wavepacket of the width $\sigma$ which moves with the velocity $\omega$ along the ring and the other atom is prepared in a similar wavepacket but on the opposite point on the ring and with the velocity $-\omega$. In the course of time evolution in the moving frame, such an initial product state reveals spreading of the probability density along the center of mass coordinate $X$ as expected from the free-particle form of the Hamiltonian $H_{CM}$, Eq.~(\ref{floquet_main_CM}). Along the relative distance coordinate, the probability density initially also spreads but next Anderson localization takes over, the spreading stops and the Anderson molecule forms. It is illustrated in Fig.~\ref{evolution1} where for sufficiently long time evolution the probability density in the moving frame becomes nearly stationary and shows that the atoms remain at the distance $x_1-x_2\approx \pi$ with the uncertainty $l_{loc}(\tilde E)\approx 2 k_m \tilde E/(\pi\lambda^2)\approx 0.08$ where $\tilde E\approx 1/\sigma^2$.

We have described Anderson molecules in the moving frame but the experimental detection most probably will be performed in the laboratory frame. As described in the previous paragraph, if we start with the Gaussian wavepackets and wait long enough we obtain nearly a  stationary probability density in the moving frame which can be approximated by  $|\psi(x_1,x_2)|^2\sim \exp\left[-|x_1-x_2-x_*|/l_{loc}(\tilde E)\right]$. Switching to the laboratory frame, the probability density reads 
\be
|\psi_{lab}(x_1,x_2,t)|^2\sim \exp\left(-\frac{|x_1-x_2-x_*-2\omega t|}{l_{loc}(\tilde E)}\right).
\label{psi_lab}
\ee
Keeping in mind that the atoms are on the ring of the circumference of $2\pi$, Eq.~(\ref{psi_lab}) indicates that the pattern presented in the bottom right panel of Fig.~\ref{evolution1} will be reproduced in the laboratory frame at the time moments $t=n\pi/\omega$ where $n$ is integer. If we decide to observe the atoms at different time moments, i.e. $t=t_0+n\pi/\omega$ where $t_0\ne 0$, their relative distance will be localized around $x_*+2\omega t_0$.
  
Realization and detection of an Anderson molecule in the experiment with only two atoms can be nontrivial. However, it should be much easier to perform the same experiment but with the help of two Bose-Einstein condensates (BEC) of ultra-cold atoms because many molecules can be created in a single experimental realization. To give a flavor of the experimental parameters, let us consider the following two examples. 

Assume that two BECs, each consisting of $N$ atoms (e.g. $^{39}$K atoms in the hyperfine states $|F=1,~m_F=0\rangle$ and $|F=1,~m_F=-1\rangle$), form Gaussian wavepackets of the width $\sigma=33~\mu$m which are moving on a quasi-1D ring in the opposite directions with the velocities $\pm4.1$~mm/s. The ring is realized by the toroidal trap of the radius $R=40~\mu$m and the transverse harmonic confinement of the frequency $\omega_\perp=2\pi \times 10$~kHz, corresponding to the transverse radius 15~nm \cite{Wright2000}. The inter-species s-wave scattering length can be modulated in time by changing magnetic field close to the Feshbach resonance at 113.76~G \cite{Tanzi2018}. If the scattering length is periodically modulated with frequency $\omega=2\pi \times 16$~Hz and amplitude $7.9$~nm (which corresponds to $\lambda=4.0$) and the modulation consists of $k_m=6$ harmonics with random phases, then diatomic Anderson molecules are expected to form, where the relative distance between the two atoms is determined with the uncertainty $l_{loc}\approx 14~\mu$m. The molecules should form after the time of propagation that lasts $t \approx l_{loc} (\frac{\hbar}{m\sigma})^{-1} \approx 300$~ms.

Another option to modulate the interaction strength between atoms in a quasi-1D trap is to use confinement-induced resonances. For example assume that two BECs consisting of $^{133}$Cs atoms in two different hyperfine states form two Gaussian wavepackets of the width $\sigma=8.9~\mu$m which are moving with the velocities $\pm3.6$~mm/s in a toroidal trap of the radius $R=40~\mu$m and the transverse harmonic confinement of the frequency $\omega_\perp=2\pi \times 14.5$~kHz. If the inter-species scattering length is tuned to the value $a_s\approx 1370 a_0$, for example using an appropriate magnetically-induced Feshbach resonance \cite{Chin2004}, a confinement-induced resonance takes place. The 1D coupling parameter, $\lambda_0+\lambda f(t)$, can now be modulated by changing the transverse confining frequency $\omega_\perp$ \cite{Olshanii1998,Haller2010}. When the modulation is done with frequency $\omega=2\pi \times 14$~Hz and amplitude $\lambda = 34$ (that is 20\% of the coupling parameter $\lambda_0=m R \omega_\perp a_s/(\pi\hbar) = 170$) and consists of $k_m=18$ harmonics with random phases, then diatomic Anderson molecules are expected to form, where the relative distance between the two atoms is determined with the uncertainty $l_{loc}\approx 7.1~\mu$m. The molecules should form after the time of propagation that lasts $t \approx l_{loc} (\frac{\hbar}{m\sigma})^{-1} \approx 140$~ms.

In both examples, each atom from the first BEC interacts with each atom from the other BEC. Therefore, many-body calculations are needed in order to estimate the efficiency of the creation of Anderson molecules in such {\it quantum chemical reactions}.


\section{A pair of Anderson molecules on a ring}
\label{sec2}

\begin{figure}        
\includegraphics[width=0.8\columnwidth]{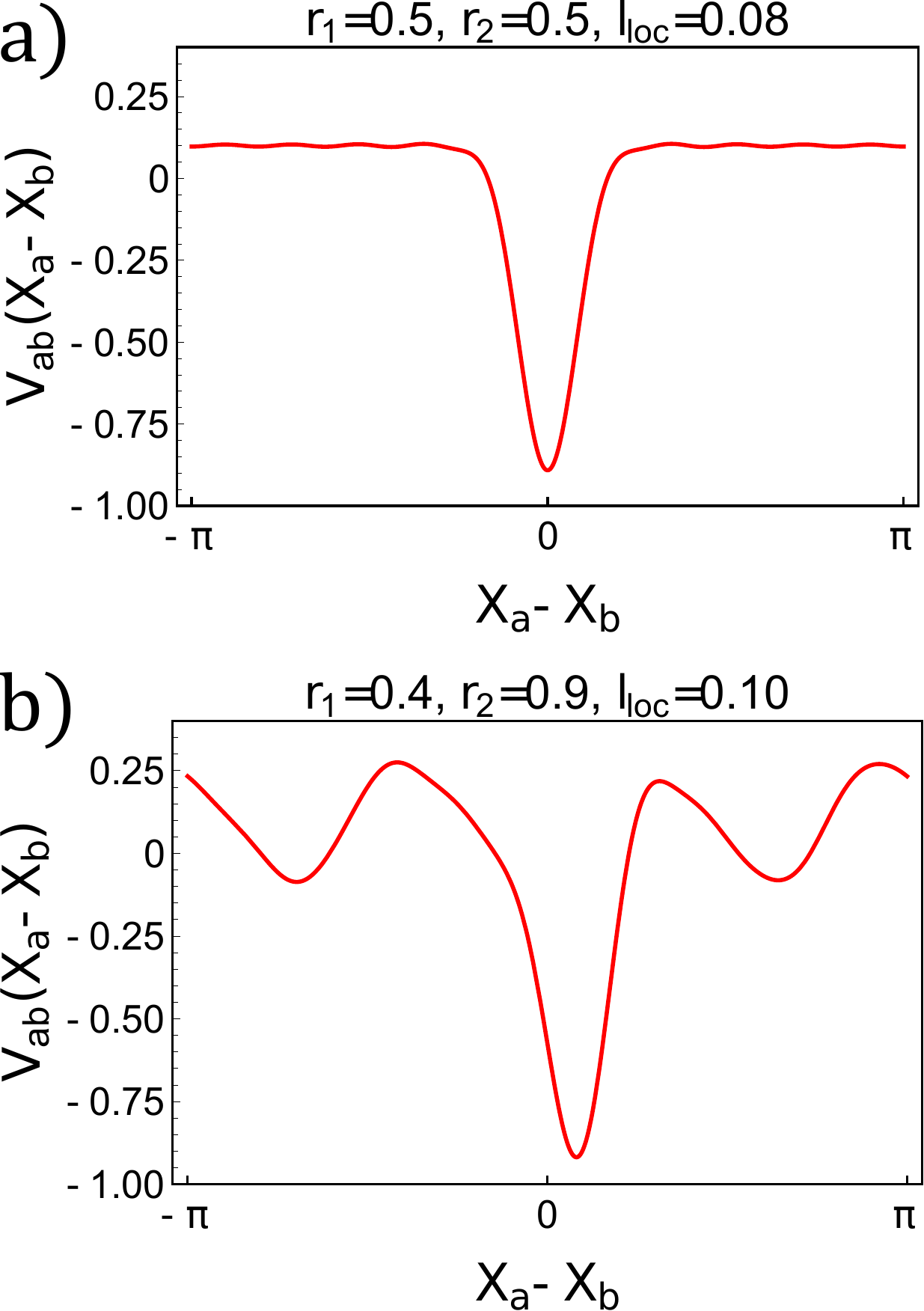}       
\caption{Interaction potentials $V_{ab}(X_a-X_b)$ between two Anderson molecules, see Eq.~(\ref{two_mol_pot}). Panel~(a): the sizes of the molecules are the same, i.e. $2r_1=2r_2=1$ with the uncertainties given by $l_{loc}(E_{a,0})=l_{loc}(E_{b,0})=0.08$. The Fourier coefficients in (\ref{two_mol_pot}) are chosen so that the interaction potential is a Gaussian well $V_{ab} \approx -\exp\left[-8(X_1-X_2)^2 \right]+\mathrm{const}$. Panel~(b): if the states of the molecules change, i.e. their sizes become $2r_1=0.8$ and $2r_2=1.8$ while the uncertainties become $l_{loc}(E_{a,0})=l_{loc}(E_{b,0})=0.10$, the interaction potential $V_{ab}$ changes too. Both in panel (a) and (b), the non-zero Fourier components in Eq.~(\ref{two_mol_pot}) are the same, i.e. $f_{3}=0.097 e^{2.1i}$, $f_{6}=0.089e^{1.1i}$, $f_{9}=0.13e^{-0.9i}$, $f_{12}=0.064e^{-0.9i}$, $f_{15}=0.049e^{-1.9i}$, $f_{18}=0.047e^{-3.0i}$, $f_{21}=0.025e^{2.4i}$, $f_{24}=0.016e^{1.4i}$, $f_{27}=0.010e^{0.4i}$, $f_{30}=0.0059e^{-0.6i}$.
}
\label{potentials}   
\end{figure} 

Increasing the number of atoms which are moving on the ring with different velocities allows one to realize more complex molecular structures and various mixtures of molecules and atoms. In this section, rather than considering all of numerous possibilities, we discuss one particular example as a proof of concept.

Let us consider four atoms with the same mass but in different hyperfine states revolving around the 1D ring with the velocities $\pm 3\omega/2$ and $\pm 7\omega/2$ as depicted in Fig.~\ref{all_setups}(b). We assume that a magnetic field is applied and it is periodically modulated in time so that the s-wave scattering lengths characterizing scattering of atoms in different hyperfine states are oscillating in time. The Hamiltonian of the system reads 
\be
H=\frac{p_1^2+p_2^2+p_3^2+p_4^2}{2}+2\pi\sum_{i<j}^4\left[\lambda_{ij}+\lambda f(t)\right]\delta(x_i-x_j),
\label{hamilt_pair}
\ee
where $f(t)$ is given in Eq.~(\ref{f_coeffs}), $\lambda_{ij}$'s are determined by the scattering lengths and for simplicity we have assumed that the amplitude $\lambda$ of the time modulation of all s-wave scattering lengths is the same. 

Applying the same secular approximation approach as in Sec.~\ref{sec1}, i.e. switching to the frame co-moving with the atoms and averaging the Hamiltonian over time (with the assumption that all quantities are slowly evolving) we obtain the following effective secular Hamiltonian
\begin{multline}
H_{\rm eff} = \frac{p_1^2+p_2^2+p_3^2 +p_4^2}{2} + \lambda \sum_n \left[f_{-2n}e^{in(x_1-x_2)} \right. \\ \left.+ f_{-2n}e^{in(x_3-x_4)} +f_{-3n}e^{in(x_2-x_3)} + f_{-5n}e^{in(x_1-x_3)} \right. \\ \left. + f_{-5n}e^{in(x_2-x_4)} + f_{-7n}e^{in(x_1-x_4)} \right]. 
\label{hamil_floquet4_2}
\end{multline}

Let us start with the case when in Eq.~(\ref{f_coeffs}) the Fourier coefficients $f_k$ with $k$ divisible by 3, 5 or 7 vanish. Then, the effective Hamiltonian (\ref{hamil_floquet4_2}) reduces to 
\bea
H_{{\rm eff},0} &=& \frac{p_1^2+p_2^2}{2}  + \lambda \sum_n f_{-2n}e^{in(x_1-x_2)}  \cr
&&+\frac{p_3^2 +p_4^2}{2}+ \lambda \sum_n f_{-2n}e^{in(x_3-x_4)},
\label{hamil_floquet4_2_0}
\eea
which is the sum two Hamiltonians of the form of (\ref{floquet_2atoms}) which is analyzed in Sec.~\ref{sec1}. If the Fourier coefficients in (\ref{hamil_floquet4_2_0}) are chosen randomly, two Anderson molecules can form. The first molecule is described by the center of mass position $X_a=(x_1 + x_2)/2$ and momentum $P_a=p_1 + p_2 $ and the other one by $X_b=(x_3 + x_4)/2$ and $P_b=p_3 + p_4$. Assume that the energies of the molecules at rest (i.e. for $P_{a,b}=0$) are $E_{a,0}$ and $E_{b,0}$, their average sizes are $\la x_1-x_2\ra \approx 2 r_a$ and $\la x_3-x_4\ra \approx 2r_b$, and the uncertainties of the sizes are determined by the localization lengths $l_a = l_{loc}(E_{a,0})$ and $l_b = l_{loc}(E_{b,0})$.

Having two Anderson molecules formed we can now modify modulation of the s-wave scattering lengths by turning on the Fourier harmonics in Eq.~(\ref{f_coeffs}) with $k$ divisible by 3, 5 and 7. We assume that the modulus of the corresponding coefficients $f_k$ are much smaller than the modulus of the coefficients in (\ref{hamil_floquet4_2_0}). Then, in the center of mass coordinates, the entire effective Hamiltonian reads
\be
H_{\rm eff} = E_{a,0} + E_{b,0} + \frac{P_a^2 + P_b^2}{4}+V_{ab}(X_a-X_b),
\ee
with
\bea
V_{ab}(X_a-X_b)  &\approx & \lambda \sum_n \left[f_{-3n}e^{-in(r_a+r_b)}  +f_{-5n}e^{in(r_a-r_b)}
\right.
\cr
&& 
\left. + f_{-5n}e^{-in(r_a-r_b)} + f_{-7n}e^{in(r_a+r_b)}\right] 
\cr
&& \times \frac{e^{in(X_a-X_b)}}{\left(1+l_a^2n^2/4\right)\left(1+l_b^2n^2/4\right)}.
\label{two_mol_pot}
\eea
Equation~(\ref{two_mol_pot}) describes the interaction potential between the Anderson molecules which  depends on their internal states. Interestingly, the shape of the potential $V_{ab}$ can be engineered. Indeed, a proper choice of the Fourier coefficients $f_k$ in (\ref{two_mol_pot}) allows one to choose how the Anderson molecules interact with each other. Figure~\ref{potentials} shows examples of $V_{ab}$ where the Fourier coefficients $f_k$ are chosen so that the interaction potential is a Gaussian well if the sizes and the uncertainties of the sizes of the molecules are the same.

Spectrum of an Anderson molecule consists of nearly degenerate eigenstates corresponding to the same localization length $l_{loc}$ but different average distances between atoms that form a molecule. If we increase energy of a molecule, then the localization length increases too and we deal with a molecule where the uncertainty of the distance between atoms is greater. Collisions of two Anderson molecules can lead to a change of their internal states.


\section{Anderson molecule in a potential well}
\label{sec3}

So far we have analyzed 1D systems with the ring geometry. Now we shall consider a different system that might be more easily realizable in the laboratory, especially in its three-dimensional version, see Sec.~\ref{sec4}. That is, we consider two distinguishable atoms moving in an infinite potential well, see Fig.~\ref{all_setups}(c). We use the units where the size of the potential well is $\pi$ and assume that initial velocities of both particles are close to a value which we denote by $\omega$. The Hamiltonian of the system is like in Eq.~(\ref{hamilt_ring_lab}) but now there are different boundary conditions because the wavefunction of the atoms has to vanish at the boundaries of the potential well.

First, let us analyze the system within classical mechanics. It is convenient to switch from the original Cartesian ($x_i$, $p_i$) variables to new momenta $J_i$ (called actions) and new position variables $\theta_i$ (angles) \cite{Lichtenberg1992}. The latter are periodic variables which change in the range $[-\pi,\pi)$. The relation between the old and new variables is very simple
\begin{equation}
J_i = |p_i|, \quad \quad x_i= |\theta_i|,
\label{semi}
\end{equation}
and results in a new form of the Hamiltonian (\ref{hamilt_ring_lab}) which now reads (we neglect $\lambda_0$, cf. Eq.~(\ref{hamilt_ring_lab}), because it does not appear in the final form of the effective Hamiltonian)
\begin{equation}
H = \frac{J_1^2 + J_2^2}{2} + 2\pi\lambda f(t)\left[\delta(\theta_1-\theta_2)+\delta(\theta_1+\theta_2)\right].
\label{hamilt_theta}
\end{equation}
One can easily check that in the absence of interactions, i.e. for $\lambda=0$, solutions of the classical equations of motion are the following: $J_i(t)=\rm constant $ and $\theta_i(t) = J_it+\theta_i(0)$ where $J_i$'s are frequencies of the periodic motion of the particles in the potential well. 

In order to analyze the system in the presence of the interactions, let us switch to the moving frame of reference i.e.
\bea
\Theta_1 &=& \theta_1-\omega t, \quad \quad I_1 = J_1-\omega, \cr
\Theta_2 &=& \theta_2-\omega t, \quad \quad I_2 = J_2-\omega,
\label{classical_moving}
\eea
where $\omega$ is the frequency of the periodic modulation of the atomic s-wave scattering length, cf. Eq.~(\ref{f_coeffs}). We are interested in motion of the atoms with velocities close to $\omega$ which corresponds to $I_i\approx 0$ and implies that for the weak interactions \cite{footnote}, $I_i$ and $\Theta_i$ are slowly varying quantities, see Appendix~\ref{appendixA}. It is worth noting that in the situation considered here both atoms move in the same direction at $t=0$, in contrast to the example discussed in Sec.~\ref{sec1}. Keeping $I_i$ and $\Theta_i$ constant and averaging the Hamiltonian over time we obtain the following classical secular Hamiltonian (\ref{hamilt_ring_lab}), 
\begin{equation}
H_{\rm eff} = \frac{I_1^2+I_2^2}{2} + \lambda\sum_{n=-k_m/2}^{k_m/2} f_{-2n}e^{in(\Theta_1+\Theta_2)}.
\label{heff_theta}
\end{equation}
If we substitute $\Theta_1 \rightarrow x_1$, $\Theta_2\rightarrow -x_2$, $I_1\rightarrow p_1$ and  $I_2 \rightarrow -p_2$ we obtain the Hamiltonian of the same form as in Eq.~(\ref{floquet_2atoms}). Thus, the system behaves in the same way as two atoms on a ring and if we quantize the classical Hamiltonian (\ref{heff_theta}) and choose $f(t)$ so that the effective potential in (\ref{heff_theta}) is a disordered potential, we can realize the same kind of Anderson molecules as described in Sec.~\ref{sec1}.

\begin{figure}[t]	            
\includegraphics[width=1.\columnwidth]{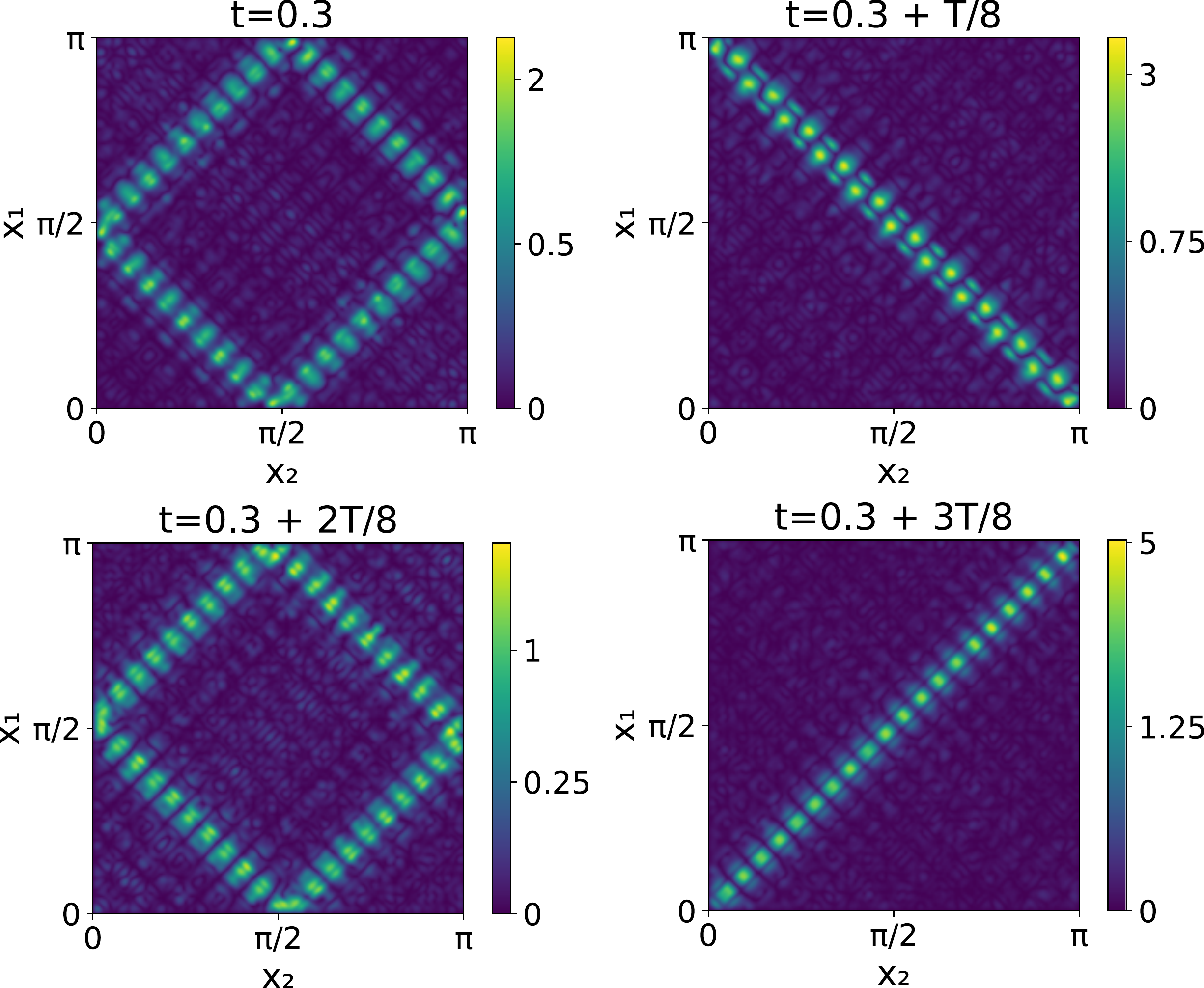}       
\caption{Time evolution, in the laboratory frame, of the modulus of the wavefunction $|\psi(x_1, x_2, t)|$ of two atoms prepared initially in a Gaussian state with the standard deviation $\sigma = 0.05$ and centered at $x_1 = x_2 = \pi/4$.  [Numbers in the color bars correspond to the values of the probability density $|\psi(x_1, x_2, t)|^2$.] At $t=0$ both wavepackets move in the same direction with velocity $\omega$. In the moving frame (\ref{1Dbox_moving_frame}) the atoms are described by the Hamiltonian (\ref{quant_secular}) which possesses the same form as the Hamiltonian (\ref{floquet_main}) and for the chosen parameters (i.e. $\lambda=600$, $f_k=e^{i\phi_k}/\sqrt{k_m}$ in Eq.~(\ref{f_coeffs}) with $k_m = 120$ and random phases $\phi_k$) an Anderson molecule is formed. How such a molecule looks in the laboratory frame is shown in the present figure for different moments of time within a half of the period $T=2\pi/\omega$ as indicated in the panels. At $t=0.3$ the time evolution of the probability density is already stabilized and reveals periodic behavior which is illustrated in the panels. Note that in the laboratory frame the signatures of the formation of an Anderson molecule for two atoms moving in the potential well are different than in the case when the atoms move on a ring, cf. Fig.~\ref{evolution1} and the discussion in Sec.~\ref{sec1}.
The results are valid for $\omega\gg 10^4$. 
}
\label{evolution3}   
\end{figure}

To describe the quantum behavior of the system one may either quantize the classical secular Hamiltonian (\ref{heff_theta}) or apply the fully quantum secular approximation approach described in Appendix~\ref{appendixB}. We present also the latter option because it allows us to easily show how Anderson molecules look like when they are observed in the laboratory frame in the Cartesian coordinates. Let us start with the original Hamiltonian (\ref{hamilt_ring_lab}). Eigenstates of two non-interacting atoms in the potential well, i.e. eigenstates of $H_0 = (p_1^2+p_2^2)/2$, are 
\begin{equation}
\la x_1, x_2|n_1, n_2 \rangle = \frac{2}{\pi} \sin(n_1 x_1)\sin(n_2 x_2),
\label{1Dbox_basis}
\end{equation}
with the corresponding eigenenergies $E_{n_1n_2} =(n_1^2+n_2^2)/2$. The energy levels are degenerated because $|n_1, n_2\rangle$ and $|n_2, n_1\rangle$ correspond to the same eigenvalue $E_{n_1n_2}$. Let us analyze how the operator of the contact interaction potential acts on the antisymmetric combination of the eigenstates. It turns out that  
\be
\delta(x_1-x_2)\left(\la x_1, x_2|n_1,n_2\rangle - \la x_1, x_2|n_2,n_1\rangle\right) = 0, 
\ee
and thus to describe the interactions between two atoms we may restrict to the symmetric subspace,
\be
|n_1,n_2\rangle_S = \left\{
\begin{matrix}
\left(|n_1,n_2\rangle + |n_2,n_1\rangle\right)/\sqrt{2}, & n_1<n_2, \\
|n_1,n_1\rangle, \hfill & n_1=n_2.
\end{matrix}
\right.
\label{sym_basis}
\ee
In order to obtain the quantum secular Hamiltonian we perform the time dependent unitary transformation,
\begin{equation}
|\bar{n}_1, \bar{n}_2\rangle\equiv e^{-i(n_1+n_2)\omega t}|n_1,n_2\rangle_S,
\label{1Dbox_moving_frame}
\end{equation}
(which is a quantum counterpart of the classical transformation (\ref{classical_moving}) to the moving frame) and average the resulting Hamiltonian over time which yields matrix elements of the secular Hamiltonian
\bea
\langle\bar{n}'_{cm},\bar{n}'| H_{\rm eff} |\bar{n}_{cm}, \bar{n}\rangle &=& \left[\left(\frac{n_{cm}^2}{4}+n^2\right)\delta_{n - n'} \right.
\cr &&
\left. + 2\lambda f_{2(n-n')}\right]\delta_{n_{cm}-n_{cm}'}, 
\cr &&
\label{quant_secular}
\eea
where $n_{cm} = n_2-n_1$ and $n = (n_1 +n_2)/2-\omega$. The effective Hamiltonian (\ref{quant_secular}) accurately reproduces the exact behavior of the system in the resonant Hilbert subspace, i.e. the subspace spanned by the states $| \bar{n}_1, \bar{n}_2\rangle$ for which the resonant conditions $E_{n_1+1,n_2}-E_{n_1,n_2} \approx \omega$ and $E_{n_1,n_2+1}-E_{n_1,n_2} \approx \omega$ (or equivalently $n_1 \approx \omega$ and $n_2 \approx \omega$) are fulfilled, see Appendix~\ref{appendixB}. 

Note that the formula (\ref{quant_secular}) for the matrix elements of the secular Hamiltonian for two atoms in the potential well is the same as in the case of two atoms on a ring. It is apparent when we substitute $n_{cm} \rightarrow P$, $n \rightarrow p$ and $2 \lambda \rightarrow \lambda$ and compare (\ref{quant_secular}) with (\ref{floquet_main}). The additional factor of 2 in the latter substitution is related to the fact that in the present case there is a restriction to the symmetric states where $n_1\leq n_2$ or equivalently $n_{cm}\geq 0$. 

Two atoms in the potential well behave in the same way as two atoms on a ring and can reveal the same Anderson localization phenomena. However, the basis states in the present case are given by different expressions [see Eq.~(\ref{1Dbox_basis})] than in Sec.~\ref{sec1} and therefore the eigenstates look somewhat different when plotted in the Cartesian coordinates $x_1$ and $x_2$. In the case of two atoms on a ring, when we observe an Anderson molecule in the laboratory frame, a wavefunction is always localized along $x_1-x_2$ but the localization point changes periodically in time, see Eq.~(\ref{psi_lab}). In the case of two atoms in the potential well, the eigenstates reveal in the laboratory frame periodic oscillations between Anderson localization of the relative position $x_1-x_2$ and the localization of the center of mass position $(x_1+x_2)/2$. Figure~\ref{evolution3} shows signatures of the formation of an Anderson molecule which can be observed in the laboratory frame.


\section{Three-dimensional Anderson molecules}
\label{sec4}

Anderson molecules can be also realized in 3D space. Two atoms moving on a 3D torus or in a 3D potential well can form an Anderson molecule if the strength of the contact interactions between atoms is properly modulated in time. A 3D torus is a mathematical model only and cannot be realized in the laboratory. However, two atoms in a 3D potential well, i.e. a 3D counterpart of the problem that we have analyzed in Sec~\ref{sec3}, is attainable experimentally and in the following we will focus on such a system.

A crucial difference in the character of Anderson localization in three-dimensions as compared to 1D and 2D cases is the presence of a localized-delocalized transition. According to the general theory of Anderson localization in order to observe the localization in 3D space, a disordered potential must exceed a certain critical strength \cite{Abrahams:Scaling:PRL79,MuellerDelande:Houches:2009}. The localized-delocalized transition can be also observed for fixed disorder by changing the energy of a particle. Indeed, there is the so-called mobility edge, i.e. a value of the energy that separates parts of the spectrum with localized and delocalized eigenstates. In the present section we show how to realize diatomic Anderson molecules in 3D where the localized-delocalized transition is related to dissociation of the molecule and the difference between the mobility edge and the energy of the system corresponds to the bond energy.

We consider two atoms of the same kind but in different hyperfine states which are trapped in a 3D cubic potential well, see Fig \ref{all_setups}(d). We use the units $\hbar^2\pi^2/mL^2$ and $L/\pi$ for energy and length, respectively, where $L$ is the length of the edge of the cubic well. The Hamiltonian of the system reads
\begin{equation}
H = \frac{\vect{p}_1^2+\vect{p}_2^2}{2} + 2 (2\pi)^3\left[\lambda_0+\lambda f_1(t)f_2(t)f_3(t)\right] \delta(\vect{r}_1-\vect{r}_2),
\label{Hfull3D}
\end{equation}
with $\lambda_0=a_s/(4\pi L)$ where $a_s$ is the atomic s-wave scattering length. The amplitude of the time modulation of the scattering length is characterized by $\lambda$ and 
\begin{equation}
f_j(t) = \sum_{k} f_k^{(j)} e^{ik\omega_jt},
\end{equation}
with $f_0^{(j)}=0$ and $f_k^{(j)}=f_{-k}^{(j)*}$ being independent random numbers. We assume that the frequencies $\omega_1$, $\omega_2$ and $\omega_3$ are different and their ratios are irrational numbers. At $t=0$ both atoms are supposed to move with velocities whose components along the $x$, $y$ and $z$ directions are close to the values $\omega_1$, $\omega_2$ and $\omega_3$, respectively.

The easiest way to describe the formation of 3D Anderson molecules is to derive a classical secular Hamiltonian of the system and then switch to its quantized version, cf. Sec.~\ref{sec3}. We introduce the action-angle variables ($J_{x,i}$, $\theta_{x,i}$), ($J_{y,i}$, $\theta_{y,i}$) and ($J_{z,i}$, $\theta_{z,i}$) similarly like in Eq.~(\ref{semi}) but with ($p_i$, $x_i$) replaced by ($p_{x,i}$, $x_i$), ($p_{y,i}$, $y_i$) and ($p_{z,i}$, $z_i$), where $i=1,2$ labels the atoms, and the Hamiltonian (\ref{Hfull3D}) takes the form
\bea
H &=& \frac{\vect{J}_1^2+\vect{J}_2^2}{2} + 2 (2\pi)^3[\lambda_0+\lambda f_1(t)f_2(t)f_3(t)]  \cr &&
\times \left[\delta(\theta_{x,1}-\theta_{x,2})+\delta(\theta_{x,1}+\theta_{x,2})\right] \cr &&
\times \left[\delta(\theta_{y,1}-\theta_{y,2})+\delta(\theta_{y,1}+\theta_{y,2})\right] \cr && 
\times\left[\delta(\theta_{z,1}-\theta_{z,2})+\delta(\theta_{z,1}+\theta_{z,2})\right].
\eea
In the frame moving with the atoms, which is defined by
\bea
\Theta_{x,i} = \theta_{x,i}-\omega_1 t, \quad\quad  I_{x, i} = J_{x, i}-\omega_1, \cr
\Theta_{y,i} = \theta_{y,i}-\omega_2 t, \quad\quad  I_{y, i} = J_{y, i}-\omega_2, \cr
\Theta_{z,i} = \theta_{z,i}-\omega_3 t, \quad\quad  I_{z, i} = J_{z, i}-\omega_3,
\eea
all dynamical quantities vary slowly and averaging the Hamiltonian over time yields
\bea
H_{\rm eff} &=& \frac{\vect{I}_1^2+\vect{I}_2^2}{2} + 2 \lambda V_1(\Theta_{x,1}+\Theta_{x,2}) \cr &&\times V_2(\Theta_{y,1}+\Theta_{y,2})
V_3(\Theta_{z,1}+\Theta_{z,2}),
\label{Hsec3D}
\eea
where a constant term is omitted and
\begin{equation}
V_j(\Theta) = \sum_k f_{-2k}^{(j)} e^{ik\Theta}.
\label{Vi}
\end{equation}
 As an example we will focus on the Fourier coefficients 
\begin{equation}
f_{2k}^{(j)} = \frac{1}{\sqrt{k_m}\pi^{1/4}}e^{-k^2/(2k_m^2)}e^{i\phi_{2k}^{(j)}},
\label{Fourier3D}
\end{equation}
where $\phi_{2k}^{(j)}$'s are random numbers chosen from a uniform distribution in the interval $[0,2\pi)$.
Defining the variables 
\bea
\vect{r} = \vect{\Theta}_1+\vect{\Theta}_2, \quad\quad \vect{p} = \frac{\vect{I}_1+\vect{I}_2}{2}, \\
\vect{R} = \vect{\Theta}_1-\vect{\Theta}_2, \quad\quad\vect{P} = \frac{\vect{I}_1-\vect{I}_2}{2},
\eea
we obtain the secular Hamiltonian in the final form 
\bea
H_{\rm eff} = 
 \vect P^2 +2 \left[\frac{\vect p^2}{2} + V(\vect r)\right],
\label{3Deff}
\eea
where the effective potential
\be
V(\vect r)=\lambda V_1(x)V_2(y)V_3(z).
\label{Vofr}
\ee
In order to find the range of the parameters where the first order secular Hamiltonian (\ref{3Deff}) is valid, one can calculate the second order terms which turn out to be proportional to 
\be
\frac{\lambda^2}{(k\omega_1+m\omega_2+n\omega_3)^2}\;e^{-(k^2+m^2+n^2)/4k_m^2},
\label{secondorder}
\ee
where $k$, $m$ and $n$ are non-zero integers. The ratios of the frequencies $\omega_j$ are irrational numbers but it may still happen that the denominator in (\ref{secondorder}) is close to zero and then the second order corrections may not be neglected. To avoid such a small denominator problem, we assume $\omega_1>2k_m(\omega_2+\omega_3)$ which ensures that even if the denominator is close to zero, the second order terms are suppressed by the exponential function and the first order secular Hamiltonian (\ref{3Deff}) is a valid description of the system. Having the classical secular Hamiltonian we perform its quantization by defining the operators $\vect P^2=-\nabla_{\vect R}^2$ and $\vect p^2=-\nabla_{\vect r}^2$ in the Hilbert space spanned by 3D generalization of the eigenstates (\ref{1Dbox_basis}) of two non-interacting particles in a potential well.

The quantum version of the Hamiltonian (\ref{3Deff}) has the form of the Hamiltonian of two quantum particles whose center of mass position $\vect R$ is described by the kinetic energy term $\vect P^2$ while in the space of the relative position $\vect r$ there is the potential (\ref{Vofr}) which is a product of three independent disordered potentials (\ref{Vi}). An Anderson molecule forms if the wavefunction of the system is exponentially localized in the space of the relative position of the particles due to the presence of the disordered potential $V(\vect r)$. 

A system described by the Hamiltonian in the bracket in Eq.~(\ref{3Deff}) was analyzed in Ref.~\cite{delande17} by means of the transfer matrix method. For relatively large $k_m$ in Eq.~(\ref{Fourier3D}), values of the disordered potential $V(\vect r)$ have a Gaussian distribution with zero mean. If we assume that $\vect r$ is not limited to the finite space of the 3D cubic box, we obtain that the correlation function of the potential drops to zero like 
\be
\overline{V(\vect r')V(\vect r'+\vect r)}=\lambda^2\; e^{-r^2/2\xi^2},
\ee
where the overbar denotes the averaging over the disorder realizations and $\xi=\sqrt{2}/k_m$ is the correlation length. There are three energy scales in the Anderson localization problem considered here: the energy $E$ of the relative degree of freedom of two particles, the strength of the disordered potential $\lambda$ and the correlation energy $E_\xi=1/\xi^2$. We choose $E_\xi$ as the natural energy scale, then the ratio of the mobility edge $E_c$ and $E_\xi$ depends only on $\lambda/E_\xi$ \cite{Kuhn:Speckle:NJP07}. Assuming that the strength of the disorder potential is $\lambda=E_\xi$, an Anderson molecule can form if the eigenenergy $E/E_\xi$ of the relative position degree of freedom of the particles is smaller than the mobility edge $E_c/E_\xi\approx 0.064\pm 0.004$ \cite{delande17}. Then, the corresponding wavefunction $\psi_E(\vect r)\sim\exp[-|\vect r-\vect r_*|/l_{loc}(E)]$ and the size of an Anderson molecule is given by $r_*$ with the uncertainty determined by the localization length $l_{loc}(E)$. When the energy $E/E_\xi$ approaches the mobility edge $E_c/E_\xi$, the localization length $l_{loc}(E)/\xi$ diverges signaling the localized-delocalized transition which corresponds to dissociation of an Anderson molecule. The bond energy of a molecule corresponds to $E_c-E$.

In order to realize Anderson molecules in a 3D cubic potential well, the localization length must fulfill $l_{loc}(E)\ll \pi$ and it seems impossible to observe dissociation of the molecules due to the localized-delocalized Anderson transition. However, by choosing sufficiently large $k_m$ in (\ref{Fourier3D}) we can always choose the correlation length $\xi$ so small that $l_{loc}(E)/\xi$ is as large as we wish but $l_{loc}(E)\ll \pi$. In other words, with the help of the spatially finite system it is possible to investigate the vicinity of the localized-delocalized transition arbitrarily close to the critical point.  

Two particles described by the Hamiltonian (\ref{3Deff}) can form an Anderson molecule and let us now analyze how signatures of its formation look like in the laboratory frame when we measure positions of the atoms in the Cartesian coordinates which we denote by $\vect r_1$ and $\vect r_2$. Assume that the system is prepared in a state $\psi(\vect r_1,\vect r_2,t)$ which corresponds to an eigenstate of the Hamiltonian (\ref{3Deff}) with the eigenenergy below the mobility edge. It means that an Anderson molecule is formed. If we plot the reduced probability density in the space spanned by $x_1$ coordinate of the first atom and $x_2$ coordinate of the other one, i.e.
\be
\rho(x_1,x_2,t)=\int d^3\vect r_1'd^3\vect r_2' |\psi(\vect r_1',\vect r_2',t)|^2\delta(x_1-x_1')\delta(x_2-x_2'),
\ee
we will observe similar behavior like in Fig.~\ref{evolution3}. That is, within every period $2\pi/\omega_1$ the probability density $\rho(x_1,x_2,t)$ will reveal oscillations between Anderson localization along the relative position around $x_1-x_2\approx 0$ and along the center of mass position around $x_1+x_2\approx \pi$. Similar behavior will be observed if we calculate the reduced probability density in the $y_1y_2$ or $z_1z_2$ spaces but with the period $2\pi/\omega_2$ and $2\pi/\omega_3$, respectively. Because the ratios of the frequencies $\omega_j$ are irrational numbers, it is possible to find a moment of time $t$ when the full probability density $|\psi(\vect r_1,\vect r_2,t)|^2$ is localized around $\vect r_1-\vect r_2\approx 0$.


\section{Summary and Perspectives}
\label{conclusions}

If the interaction potential between two atoms changes in a disordered way with their relative distance, they can form molecules of a completely different nature than ordinary molecules. That is, it is not attractive interactions between atoms that are responsible for the formation of the molecules, but it is the destructive interference phenomena and the resulting Anderson localization that lead to the formation of bound states which we dub Anderson molecules. 

Even though there are no disordered interaction potentials in nature, we show that proper time modulation of the strength of the original interaction potentials between atoms creates effective interactions which allow for the realization of Anderson molecules. Such effective interactions are the result of resonant couplings between different harmonics of the translation motion of atoms and the time-periodic modulation of the original atom-atom interactions. It is possible to engineer various effective interaction potentials and to create different molecular structures and different mixtures of interacting molecules and atoms. In the present paper we show how to realize diatomic Anderson molecules for atoms moving on a 1D ring or in a 1D potential well and also how to create Anderson molecules in 3D space. As an example of more complex molecular structures we show that two diatomic Anderson molecules can be realized and the interaction potential between them can be engineered at will. 

Experimentally it should be possible to realize Anderson molecules in ultra-cold atomic gases. For example if two Bose-Einstein condensates, that consist of different atomic species, move periodically in a toroidal trap in the opposite directions, Anderson molecules can be produced if the inter-species s-wave scattering length is properly modulated in time. The latter can be done by applying a magnetic field which oscillates close the value corresponding to the inter-species Feshbach resonance or using confinement-induced resonances. One can substitute a toroidal trap by any trapping potential provided it is not a harmonic potential, because periodic motion of an atom in such a potential possesses only one harmonic, while many harmonics are needed to create disordered effective interactions.

There are several possible directions for further research. We have concentrated on diatomic Anderson molecules but it should be possible to create Anderson molecules consisting of a larger number of atoms. It is also interesting whether a strongly interacting many-body system can reveal many-body localization if the strength of the interactions between particles is modulated in time in a disordered way. Originally many-body localization has been considered in systems with strong spatial disorder which prevents thermalization because of the existence of local integrals of motion, which carry information about the initial state of a system  \cite{Basko06,Oganesyan07,Znidaric08,Huse14,Rahul15,Ponte2015,Zakrzewski16,Bar2016,Abanin2019}. Temporal disorder has been proposed to create effective external disordered potentials for atoms \cite{Mierzejewski2017}. The present work indicates that effective interaction potentials between particles can change in a disordered way with the relative distances between particles if the strengths of the original interactions are properly modulated in time. Many-body localization in this case should be related to the formation of clusters of particles which can move in space as a whole \cite{Mondaini2017}. It should be also mentioned that randomly fluctuating external force can be used to modify effective interactions between two solitons in non-linear dissipative media. The dissipation rate determines positions of regularly distributed minima of the effective potential which can host 
bound states of two solitons \cite{Malomed1995}.


\section*{Acknowledgements}

We are grateful to Peter Hannaford for discussion and valuable comments. 
Support of the National Science Centre Poland via Projects No. 2016/20/W/ST4/00314 and 2019/32/T/ST2/00413 (K.G.) and 2018/31/B/ST2/00349 (K.S.) is gratefully acknowledged. K.G. acknowledges the support of the Foundation for Polish Science (FNP).


\appendix
\section{Classical secular approximation approach}
\label{appendixA}

Let us consider a classical particle described by the time-periodic Hamiltonian 
\be
H(t+T)=H(t)=H_0(x,p)+\lambda V(x,t),
\label{appenHcl}
\ee
where $H_0$ is the unperturbed part of the Hamiltonian and a particle, in the absence of the perturbation (i.e. for $\lambda=0$), can perform periodic motion. In order to describe the system when a particle is resonantly perturbed (i.e. when the driving period $T=2\pi/\omega$ is close to the period of an unperturbed particle trajectory) let us perform a canonical transformation from the Cartesian coordinates $p$ and $x$ to the so-called action-angle variables $J$ and $\theta\in [0,2\pi)$ \cite{Lichtenberg1992,Buchleitner2002}. Then, $H=H_0(J)+\lambda V(\theta,J,t)$ and for $\lambda=0$, solutions of the classical equations of motion read $J(t)=\rm constant$ and $\theta(t)=\Omega(J)t+\theta(0)$ where $\Omega(J)=dH_0(J)/dJ$ is the frequency of an unperturbed periodic orbit. When the perturbation is turned on, we choose the initial action for a particle $J\approx J_0$ where $J_0$ fulfills the resonant condition $\Omega(J_0)= \omega$. To obtain an effective Hamiltonian which describes such resonant dynamics of a particle it is useful to switch to the moving frame 
\be
\Theta=\theta-\omega t, \quad \quad  I=J.
\label{appen_class_mov}
\ee 
It results in a new form of the Hamiltonian $H=H_0(I)-\omega I+\lambda V(\Theta+\omega t,I,t)$. Due to the resonance condition, $\Omega(J_0)=\omega$, both $I$ and $\Theta$ vary slowly if initially $I\approx J_0$ and the time-periodic perturbation is weak, i.e.
\bea
\dot I&=&-\frac{\partial H}{\partial \Theta}={\cal O}(\lambda), \\
\dot \Theta&=&\frac{\partial H}{\partial I}=\Omega(I)-\omega+{\cal O}(\lambda)\approx {\cal O}(\lambda).
\eea
It allows us to obtain the effective Hamiltonian by keeping $I$ and $\Theta$ fixed and averaging the exact Hamiltonian over time,
\be
H_{\rm eff}=H_0(I)-\omega I+\frac{\lambda}{T}\int_0^T dt V(\Theta+\omega t,J_0,t).
\ee
The Hamiltonian $H_{\rm eff}$ is the result of the first order secular approximation and it describes behavior of a particle in the vicinity of the resonant trajectory for weak time-periodic perturbation \cite{Lichtenberg1992,Buchleitner2002}.

\section{Qunatum secular approximation approach}
\label{appendixB}

Let us consider a quantum counterpart of the classical Hamiltonian (\ref{appenHcl}). If the canonical transformation between the Cartesian variables and the action-angle variables is linear, then the quantum secular Hamiltonian can be obtained in the same way as the classical one. In the present paper it is the case for particles moving on a ring because the Cartesian momenta are actually the action variables and the positions of particles on a ring are the angle variables. For particles in a potential well it is not longer true and one can either derive a classical secular Hamiltonian and then quantize it or perform the quantum secular approximation approach which we are going to describe here \cite{Berman1977,Giergiel2018a}. 

Let us denote eigenstates of the unperturbed Hamiltonian by $|n\ra$ where $H_0|n\ra =E_n|n\ra$, and perform the time-dependent unitary transformation to the moving frame, 
\be
{\cal U}|n\ra =e^{in\omega t}|n\ra, 
\label{appen_U_mov}
\ee
which is a quantum counterpart of the classical canonical transformation (\ref{appen_class_mov}).
It results in a new Hamiltonian $H\rightarrow {\cal U}H{\cal U}^\dagger+i(\partial_t{\cal U}){\cal U}^\dagger$ whose matrix elements read
\bea
\la n'|H|n\ra&=&(E_n-n\omega)\delta_{n'n}+\lambda\la n'|V(t)|n\ra e^{i(n'-n)\omega t}. \cr &&
\label{appenMatrix1}
\eea
Matrix elements of the first order quantum secular Hamiltonian can be obtained by averaging (\ref{appenMatrix1}) over time,
\bea
\la n'|H_{\rm eff}|n\ra&=&(E_n-n\omega)\delta_{n'n} \cr &&
+\frac{\lambda}{T}\int_0^T dt\la n'|V(t)|n\ra e^{i(n'-n)\omega t}.
\label{appenQsecularH}
\eea
If the time-periodic perturbation is weak, the secular Hamiltonian (\ref{appenQsecularH}) accurately reproduces the exact behavior of the system in the resonant Hilbert subspace which is spanned by the states $|n\ra$ that fulfill the resonant condition $E_{n+1}-E_n\approx \omega$ \cite{Berman1977,Giergiel2018a}.

\section{Magnus expansion}
\label{appendixC}

For a time-periodic quantum Hamiltonian $H(t+T)=H(t)$, the Floquet theorem states that the time evolution operator $U(t,0)$ can be written in the form \cite{Shirley1965}
\begin{equation}
U(t,0) = P(t)\exp\left[-iH_{F}t\right], 
\label{u2}
\end{equation}
where $H_{F}$ is a time-independent Hermitian operator often called the Floquet Hamiltonian, while $P(t)$ is a unitary time-periodic operator which fulfills $P(t+T) = P(t)$ and $P(0) = 1$. Equation~(\ref{u2}) implies that
\begin{equation}
H_{F} = -\frac{1}{iT}\log\left[ U(T,0)\right]. 
\label{appenH_F}
\end{equation}
Usually it is not easy to calculate the right hand side of Eq.~(\ref{appenH_F}) explicitly. However, one can use the so-called Magnus expansion of $H_{F}$ in powers of $T/(2\pi)=1/\omega$. The first three terms of this expansion (see e.g. \cite{Blanes2010}) read
\bea
H_{F}^{(0)} &=& \frac{1}{T}\int\limits_{0}^{T} d t_1 H(t_1), 
\label{appen_fristMagnus}
\\
H_{F}^{(1)}  &=& \frac{1}{2Ti}\int\limits_{0}^{T} d t_1 \int\limits_{0}^{t_1} d t_2 \left[H(t_1), H(t_2) \right], \\
H_{F}^{(2)} &=& -\frac{1}{6T}\int\limits_{0}^{T} d t_1 \int\limits_{0}^{t_1} d t_2 \int\limits_{0}^{t_2} d t_3 ([H(t_1),[ H(t_2),H(t_3)]] 
\cr && 
+[H(t_3),[ H(t_2),H(t_1)]]). 
\label{magn}
\eea
The first term (\ref{appen_fristMagnus}) is identical to the first order secular Hamiltonian (\ref{appenQsecularH}) if before calculating it we perform the time-dependent unitary transformation to the moving frame (\ref{appen_U_mov}). Analysis of the second and third terms allows one to check if the restriction to the first term is sufficient to accurately describe a system.


\begin{thebibliography}{82}
\bibitem[{foo()}]{footnote}
\bibinfo{note}{Strictly speaking, in the classical description a smooth version
  of the Dirac-delta potential has to be used, i.e. $\delta(x)\approx
  (2\pi)^{-1}\sum_{|n|<n_{max}}e^{inx}$, otherwise $\lambda\delta(x)$ would
  never be a weak perturbation.}
\end{thebibliography}

\end{document}